\documentclass{sig-alternate}

\newcommand{\cut}{{\rm cut}}
\newcommand{\Ncut}{{\rm Ncut}}
\newcommand{\s}{{\rm s}}
\newcommand{\diag}{{\rm diag}}

\newcommand{\R}{{\cal R}}
\newcommand{\tf}{{\rm tf}}
\newcommand{\df}{{\rm df}}
\newcommand{\sre}{{\rm sre}}
\newcommand{\svd}{{\rm svd}}
\newcommand{\nnz}{{\rm nnz}}

\begin{document}
%
\conferenceinfo{CIKM}{'01 November 5-10, 2001, Atlanta, Georgia. USA}
\CopyrightYear{2001} 

\title{Bipartite Graph Partitioning and Data 
Clustering\titlenote{Part of this work was done while Xiaofeng He
was a graduate research assistant at NERSC, Berkeley National Lab.}
}
%
%

\numberofauthors{3}
%

\author{
%
\alignauthor Hongyuan Zha \\[2pt] Xiaofeng He\\
       \affaddr{Dept. of Comp. Sci. \& Eng.}\\
       \affaddr{Penn State Univ.}\\
       \affaddr{State College, PA 16802}\\
       \email{\{zha,xhe\}@cse.psu.edu}
\alignauthor Chris Ding \\[2pt] Horst Simon\\
       \affaddr{NERSC Division}\\
       \affaddr{Berkeley National Lab.}\\
       \affaddr{Berkeley, CA 94720}\\
       \email{\{chqding,hdsimon\}@lbl.gov}
\alignauthor Ming Gu\\
       \affaddr{Dept. of Math.}\\
       \affaddr{U.C. Berkeley}\\
       \affaddr{Berkeley, CA 94720}\\
       \email{mgu@math.berkeley.edu}
}
\maketitle
\begin{abstract}
Many data types arising from data mining applications
can be modeled as bipartite graphs, examples include
terms and documents in a text corpus, customers and 
purchasing items in market basket analysis and reviewers
and movies in a movie recommender system. In this paper,
we propose a new data clustering method based on
partitioning the underlying bipartite graph. The partition
is constructed by minimizing a {\it normalized}
sum of edge weights between {\it unmatched} pairs of vertices
of the bipartite graph.
We show that an approximate solution to the minimization
problem can be obtained by computing
a partial singular value decomposition (SVD) 
of the associated edge weight
matrix of the bipartite graph. We  point out the connection
of our clustering algorithm to correspondence analysis used in
multivariate analysis. We also briefly discuss the issue
of assigning data objects to multiple clusters.
In the experimental results, we apply our clustering
algorithm to the problem of document clustering to illustrate its
effectiveness and efficiency.
\end{abstract}




\section{Introduction}\label{sec:int}
Cluster analysis is an important tool for exploratory data mining
applications arising from many diverse disciplines. Informally,
cluster analysis seeks to partition a given data set into compact
clusters so that data objects within a cluster are more similar
than those in distinct clusters. The literature on cluster analysis
is enormous including contributions from many research communities.
(see \cite{Everitt,Gordon} for 
recent surveys of some classical approaches.) Many traditional
clustering algorithms are based
on the assumption that the given dataset
consists of covariate information (or attributes) for each individual
data object, and cluster analysis can be cast as a problem of
grouping a set of $n$-dimensional vectors each representing
a data object in the dataset. A familiar example
is document clustering using the vector space 
model \cite{bele:00}. Here each document
is represented by an $n$-dimensional vector, and each
coordinate of the vector corresponds to a term in a vocabulary of size $n$.
 This formulation
leads to  the so-called term-document matrix $A=(a_{ij})$
for the representation of the collection of documents,
where $a_{ij}$ is the
so-called term frequency, i.e.,
 the number of times term $i$ occurs in document $j$.
In this vector
space model terms and documents are treated asymmetrically with
terms considered as the covariates or attributes of documents. It is
also possible to treat both terms and documents as first-class citizens
in a symmetric fashion, and consider $a_{ij}$ as the frequency of
co-occurrence of term $i$ and document $j$ as is done,
for example,  in probabilistic
latent semantic indexing \cite{hoff:99}.\footnote{Our clustering
algorithm computes an approximate global optimal solution while probabilistic
latent semantic indexing relies on the EM algorithm and therefore might be
prune to local minima even with the help of some annealing process.}
In this paper, we
follow this basic principle and propose a new approach 
to model terms and documents as vertices in a bipartite graph
with edges of the graph indicating the co-occurrence of terms and documents.
In addition
we can optionally
 use edge weights to indicate the frequency of this co-occurrence.
Cluster analysis for document collections
in this context is based on a very intuitive notion: documents
are grouped by topics, on one hand
documents in a topic tend to more heavily use the same subset
of terms which form a term cluster, and
on the other hand a topic usually is characterized
by a subset of terms and those documents heavily using those terms tend to
be about that particular topic. It is this interplay of terms and
documents which gives rise to what we call bi-clustering by which
terms and documents are simultaneously grouped into
{\it semantically coherent} clusters.

Within our bipartite graph model, the clustering problem can be
solved by constructing vertex graph partitions. Many criteria have been
proposed for measuring the quality of graph partitions 
of undirected graphs \cite{Chung,Shi}. In this paper, we show 
how to adapt those criteria for bipartite graph partitioning and
therefore solve the bi-clustering problem. A great variety of
objective functions have been proposed for cluster analysis without
efficient algorithms for finding the (approximate) optimal solutions.
We will show that our bipartite graph formulation naturally leads to
partial SVD problems for the underlying edge weight matrix
which admit efficient
{\it global} optimal solutions.
The rest of the paper
is organized as follows: in section \ref{se:bi}, we propose a 
new criterion for 
bipartite graph partitioning which tends to produce balanced
clusters. In section \ref{se:svd}, we show that our criterion
leads to an optimization problem that can be approximately
solved by computing a partial SVD
of the weight matrix of the bipartite graph. 
In section \ref{se:corr}, we make connection of our approximate
solution to correspondence analysis used in multivariate data
analysis. In section \ref{se:over}, we briefly
discuss how to deal with clusters with overlaps.
In section \ref{se:exp}, we 
describe experimental results on 
bi-clustering a dataset of newsgroup articles. We conclude the paper
in section \ref{se:con} and give pointers to future research.

\section{Bipartite graph partitioning}\label{se:bi}
We denote a graph by $G(V,E)$, where $V$ is the vertex set
and $E$ is the edge set of the
graph. A graph $G(V,E)$
is {\it bipartite} with two vertex
classes $X$ and $Y$ if $V = X\cup Y$ with
$X\cap Y = \emptyset$ and each edge in $E$ has
one endpoint in $X$ and one endpoint in $Y$. 
We consider weighted bipartite graph $G(X,Y,W)$ with
$W = (w_{ij})$ where $w_{ij} > 0$ denotes the weight of the
edge between vertex $i$ and $j$. We let $w_{ij}=0$ if there
is no edge between vertices $i$ and $j$. 
In the context
of document clustering, $X$ represents 
the set of  terms and $Y$ represents the set of documents, and $w_{ij}$
can be used to denote the number of times term $i$ occurs in
document $j$.
A vertex partition of $G(X,Y,W)$
denoted by $\Pi(A,B)$
is defined by  a partition of the vertex sets
$X$ and $Y$, respectively: $X=A\cup A^c$, and $Y=B\cup B^c$, where
for a set $S$, $S^c$ denotes its compliment. By convention,
we pair $A$ with $B$, and $A^c$ with $B^c$. We say that
a pair of vertices $ x \in X$ and $y \in Y$ is {\it matched} with
respect to a partition $\Pi(A,B)$ if there is an edge
between $x$ and $y$, and either $x \in A$ and $y \in B$ or
$x \in A^c$ and $y \in B^c$. For any two subsets of vertices
$ S \subset X$ and $T \subset Y$, define
\[ W(S,T) = \sum_{i \in S, j \in T} w_{ij},\]
i.e., $W(S,T)$ is the sum of the weights of edges with one
endpoint in $S$ and one endpoint in $T$. The quantity
$W(S,T)$ can be
considered as measuring the association
between the  vertex sets $S$ and 
$T$. In the context of cluster analysis
edge weight measures the similarity between data objects.
To partition data objects into
clusters, we seek a partition of $G(X,Y,W)$ such that the
association (similarity) 
between unmatched vertices is as small as possible.
One possibility is to consider for a partition $\Pi(A,B)$
the following quantity
\begin{equation}\label{eq:cut}
\begin{array}{ll} \cut(A,B) & \equiv W(A,B^c) + W(A^c, B)\\[3pt]
   & = \sum_{i \in A, j \in B^c} w_{ij} + 
     \sum_{i \in A^c, j \in B} w_{ij}.
\end{array}
\end{equation}
Intuitively, choosing $\Pi(A,B)$
to minimize $\cut(A,B)$ will give rise to a partition that
minimizes the sum of all the edge weights between unmatched
vertices. In the context of document clustering, we try to find
two document clusters $B$ and $B^c$ which have few terms in
common, and the documents in $B$ mostly use terms in $A$ and 
those in $B^c$ use terms in $A^c$.
Unfortunately, choosing a partition based entirely on
$\cut(A,B)$ tends to produce unbalanced clusters, i.e.,
the sizes of $A$ and/or $B$ or their compliments tend to be
small.
Inspired by the work in \cite{Chung,Driessche,Shi}, we propose
the following normalized variant of the edge cut in (\ref{eq:cut})
\[ \Ncut(A,B) \equiv \frac{\cut(A,B)}{W(A,Y) + W(X,B)}\]
\[          + \frac{\cut(A^c,B^c)}{W(A^c,Y) + W(X,B^c)}.\]
The intuition behind this criterion is that not only we
want a partition with small edge cut, but we also want the two
subgraphs formed between the matched vertices to be as dense
as possible. This latter requirement is partially
satisfied by introducing
the normalizing denominators in the above equation.\footnote{A
more natural criterion seems to be
\[\frac{\cut(A,B)}{W(A,B)}
          + \frac{\cut(A^c,B^c)}{W(A^c,B^c)}.\]
However, it can be shown that it will leads to an SVD 
problem with the same set of left and right singular vectors.} 
Our bi-clustering problem is now equivalent to
the following optimization problem
\[ \min_{\Pi(A,B)} \Ncut(A,B),\]
i.e., finding partitions of the vertex sets $X$ and $Y$ to
minimize the normalized cut of the bipartite graph $G(X,Y,W)$.

\section{Approximate solutions using singular vectors}\label{se:svd}
Given a bipartite graph $G(X,Y,W)$
and the associated partition $\Pi(A,B)$. Let us reorder
the vertices of $X$ and $Y$ so that vertices in $A$ and $B$ are
ordered before vertices in $A^c$ and $B^c$, respectively. The
weight matrix $W$ can be written in a block format
\begin{equation}\label{eq:w} W = \left[\begin{array}{cc}
       W_{11}&W_{12}\\
       W_{21}&W_{22}
       \end{array}\right],\end{equation}
i.e., the rows of $W_{11}$ correspond to 
the vertices in the vertex set $A$ and
the columns of $W_{11}$ correspond to 
those in $B$. Therefore
$G(A,B,W_{11})$  denotes the weighted bipartite graph
corresponding to the vertex sets $A$ and $B$. 
For any $m$-by-$n$ matrix 
$H = (h_{ij})$, define
\[ \s(H) = \sum_{i=1}^m \sum_{j=1}^n h_{ij},\]
i.e., $\s(H)$ is the sum of all the elements of $H$.
It is easy to see from the definition of $\Ncut$,
\[ \Ncut(A,B) = \frac{\s(W_{12}) + \s(W_{21})}{2\s(W_{11}) +\s(W_{12}) 
+\s(W_{21})}\]\[ +
\frac{\s(W_{12}) + \s(W_{21})}{2\s(W_{22}) +\s(W_{12}) 
+\s(W_{21})}.\]
In order to make connections to
SVD problems, we
first consider the case when $W$ is symmetric.\footnote{A different
proof for the
symmetric case was first derived in \cite{Shi}. However, our derivation
is simpler and more transparent and leads naturally to the SVD
problems for the rectangular case.}
It is easy
to see that with $W$ symmetric (denoting $\Ncut(A,A)$ by $\Ncut(A)$),
we have
\begin{equation}\label{eq:sym}
 \Ncut(A) = \frac{\s(W_{12})}{\s(W_{11})+\s(W_{12})}
+ \frac{\s(W_{12})}{\s(W_{22})+\s(W_{12})}.\end{equation}
Let $e$ be the vector
with all its elements equal to $1$. Let $D$ be the diagonal matrix
such that $We = De$. Then $(D-W)e=0$. Let $x=(x_i)$ be the vector with
\[  x_i = \left\{\begin{array}{rl}
                 1, & i \in A,\\
                -1, & i \in A^c.
                 \end{array}
           \right.
\] 
It is easy to verify that
\[ \s(W_{12}) = x^T(D-W)x/4.\]
Define
\[ p \equiv \frac{\s(W_{11})+\s(W_{12})}{\s(W_{11})+2\s(W_{12})+\s(W_{22})}
     =\frac{\s(W_{11})+\s(W_{12})}{e^TDe}.\]
Then
\[ \begin{array}{c}
\s(W_{11})+\s(W_{12}) = p e^TDe, \\[3pt]
   \s(W_{22})+\s(W_{12}) = (1-p)e^TDe,
\end{array}
\]
and
\begin{equation}\label{eq:n}
 \Ncut(A) = \frac{x^T(D-W)x}{4p(1-p)e^TDe}.
\end{equation}
Notice that  $(D-W)e=0$, then for any scalar $s$, we have
\[ (se+x)^T(D-W)(se+x)= x^T(D-W)x.\]
To cast (\ref{eq:n}) in the form of a Rayleigh quotient, 
we need to find
$s$ such that
\[ (se+x)^TD(se+x)= 4p(1-p)e^TDe.\]
Since $x^TDx = e^TDe$, it follows from the above equation that
$s = 1-2p$. Now let $y=(1-2p)e + x$, it is easy to see that
$y^TDe = ((1-2p)e + x)^TDe = 0$, and
\[  y_i = \left\{\begin{array}{rl}
                 2(1-p)>0, & i \in A,\\
                -2p<0, & i \in A^c.
                 \end{array}
           \right.
\]
Thus
\[ \min_{A} \Ncut(A) = \min \left\{ \frac{y^T(D-W)y}{y^TDy} 
\;\; | \;\; y \in S\right\},\]
where
\[ S=\{ y \;\; | \;\;
y^TDe =0, y_i 
\in \{ 2(1-p), -2p\} \}.\]
If we drop the constraints $y_i 
\in \{ 2(1-p), -2p\}$ and let
the elements of $y$ take
arbitrary continuous values, then the optimal $y$ can be approximated by
the following relaxed {\it continuous} minimization problem,
\begin{equation}\label{eq:y} \min \left\{ \frac{y^T(D-W)y}{y^TDy} 
\;\; | \;\;y^TDe =0\right\}.\end{equation}
Notice that it follows from $We = De$ that 
\[ D^{-1/2}WD^{-1/2} (D^{1/2}e) = D^{-1/2}e,\]
and therefore $D^{1/2}e$ is an eigenvector of
$D^{-1/2}WD^{-1/2}$ corresponding to the eigenvalue $1$. It
is easy to show that all the eigenvalues of $D^{-1/2}WD^{-1/2}$
have absolute value at most $1$ (See the Appendix). Thus the optimal $y$ in
(\ref{eq:y}) can be computed as $y = D^{1/2}\hat{y}$, where
$\hat{y}$ is the {\it second} largest 
eigenvector of $D^{-1/2}WD^{-1/2}$.

Now we return to the rectangular case for the weight matrix $W$,
and let $D_X$ and $D_Y$ be diagonal matrices such that
\begin{equation}\label{eq:xy} 
We = D_X e, \quad W^Te = D_Ye.
\end{equation}
Consider a partition $\Pi(A,B)$, and define
\[ u_i = \left\{\begin{array}{rl}
                 1, & i \in A\\
                -1, & i \in A^c
                 \end{array}
           \right., \quad
v_i = \left\{\begin{array}{rl}
                 1, & i \in B\\
                -1, & i \in B^c
                 \end{array}
           \right.
\]
Let $W$ have the block form as in (\ref{eq:w}), and consider the
augmented symmetric matrix\footnote{In \cite{heko:00}, the Laplacian
of $\hat{W}$ is used for partitioning a rectangular matrix
in the context of designing load-balanced matrix-vector multiplication
algorithms for parallel computation. However, the eigenvalue
problem of the Laplacian
of $\hat{W}$ does not lead to a simpler singular value problem.}
\[ \hat{W} = \left[\begin{array}{cc}
                   0 & W\\
                   W^T & 0
             \end{array}\right]
   = \left[\begin{array}{cc|cc}
                   0 & 0 & W_{11} & W_{12}\\
                   0 & 0 & W_{21} & W_{22}\\ \hline
                   W_{11}^T & W_{21}^T & 0 & 0 \\
                   W_{12}^T & W_{22}^T & 0 & 0
                      \end{array}\right].\]
If we interchange the second and third block rows and columns
of the above matrix, we obtain
\[ \left[\begin{array}{cc|cc}
                   0 & W_{11} & 0 & W_{12}\\
                   W_{11}^T & 0 & W_{21}^T & 0\\ \hline
                   0 & W_{21} & 0 & W_{22} \\
                   W_{12}^T & 0 & W_{22}^T & 0
                      \end{array}\right] \equiv
    \left[\begin{array}{cc}
                   \hat{W}_{11} & \hat{W}_{12}\\
                   \hat{W}_{12}^T & \hat{W}_{22}
             \end{array}\right],\]
and the normalized cut can be written as
\[ \Ncut(A,B) = \frac{\s(\hat{W}_{12})}{\s(\hat{W}_{11})+\s(\hat{W}_{12})}
+ \frac{\s(\hat{W}_{12})}{\s(\hat{W}_{22})+\s(\hat{W}_{12})},\]
a form that resembles the symmetric case (\ref{eq:sym}). Define
\[ q = \frac{2\s(W_{11}) +\s(W_{12}) 
+\s(W_{21})}{e^TD_Xe + e^TD_Ye}.\]
Then we have
\[ \Ncut(A,B) = \frac{-2x^TWy + x^TD_Xx + y^TD_Yy}{x^TD_Xx + y^TD_Yy}\]\[
              = 1- \frac{2x^TWy}{x^TD_Xx + y^TD_Yy},\]
where $x = (1-2p)e +u, y = (1-2p)e + v$. It is also easy to see that
\begin{equation}\label{eq:q} 
x^TD_Xe + y^TD_Ye = 0, \quad x_i, y_i \in \{ 2(1-q), -2q\}.
\end{equation}
Therefore,
\[ \min_{\Pi(A,B)} 
\Ncut(A,B)\]\[ = 1-\max_{x \neq 0, y \neq 0}
\left\{ \frac{2x^TWy}{x^TD_Xx + y^TD_Yy} 
\;\; | \;\;  x, y \; \mbox{\rm satisfy } (\ref{eq:q})\right\}.\]
Ignoring the discrete constraints on the elements of $x$ and $y$, we
have the following continuous maximization problem,
\begin{equation}\label{eq:yz}
\max_{x \neq 0, y \neq 0}
\left\{ \frac{2x^TWy}{x^TD_Xx + y^TD_Yy}\;\; | \;\; 
x^TD_Xe + y^TD_Ye = 0 \right\}.
\end{equation}
Without the constraints 
$x^TD_Xe + y^TD_Ye = 0$, the above problem is equivalent to
computing the largest singular triplet of $D_X^{-1/2} W D_Y^{-1/2}$
(see the Appendix).
From (\ref{eq:xy}), we have
\[ \begin{array}{c}
D_X^{-1/2} W D_Y^{-1/2} (D_Y^{1/2}e) = D_X^{1/2} e, \\[3pt]
(D_X^{-1/2} W D_Y^{-1/2})^T (D_X^{1/2}e) = D_Y^{1/2} e,
\end{array}
\]
and similarly to
the symmetric case, it is easy to show that all the 
singular values of $D_X^{-1/2} W D_Y^{-1/2}$
are at most $1$. Therefore, an optimal pair $\{x,y\}$ for
(\ref{eq:yz}) can be computed as
$x = D_X^{-1/2} \hat{x}$ and $y = D_Y^{-1/2} \hat{y}$,
where $\hat{x}$ and $\hat{y}$ are the {\it second}
largest left and right singular vectors of $D_X^{-1/2} W D_Y^{-1/2}$,
respectively (see the Appendix). 
With the above discussion, we can now summerize our
basic approach for bipartite graph clustering incorporating
a recursive procedure.

\bigskip

\begin{center}
\fbox{\parbox{7.7cm}{
{\sc Algorithm.} Spectral Recursive Embedding (SRE) 

Given a weighted bipartite graph $G = (X,Y,E)$
with its edge weight matrix $W$:

 \begin{enumerate}
   \item Compute $D_X$ and $D_Y$ and form the scaled weight matrix
         $\hat{W}=D_X^{-1/2} W D_Y^{-1/2}$.
   \item Compute the {\it second} largest left and right
         singular vectors of $\hat{W}$, $\hat{x}$ and $\hat{y}$.
   \item Find cut points $c_x$ and $c_y$ for $x=D_X^{-1/2}\hat{x}$
         and $y=D_Y^{-1/2}\hat{y}$, respectively.
   \item Form partitions $A=\{i \;\;| \;\;x_i \geq c_x\}$ and 
         $A^c=\{i \;\;| \;\;x_i < c_x\}$ for vertex set $X$, and
         $B=\{j \;\;| \;\;y_j \geq c_y\}$ and 
         $B^c=\{j \;\;|\;\; y_j < c_y\}$ for vertex set $Y$.
   \item Recursively partition the sub-graphs $G(A,B)$
          and $G(A^c,B^c)$ if necessary.
         
 \end{enumerate}
}}
\end{center}

\bigskip

Two basic strategies can be used for selecting the cut points
$c_x$ and $c_y$. The simplest strategy is to set $c_x=0$ and
$c_y=0$. Another more computing-intensive approach is to base
the selection on $\Ncut$: Check $N$ equally spaced splitting
points of $x$ and $y$, respectively, find the cut
points $c_x$ and $c_y$ with the smallest $\Ncut$ \cite{Shi}.

{\bf Computational complexity.} The major computational cost
of SRE is Step 2  for computing the left and right singular vectors
which can be obtained either by power method or more robustly
by Lanczos bidiagonalization process \cite[Chapter 9]{govl:96}. 
Lanczos method is an iterative process for computing
partial SVDs in 
which  each iterative step involves the computation of two matrix-vector
multiplications $\hat{W}u$ and $\hat{W}^Tv$ for some vectors
$u$ and $v$. The computational cost of these is 
roughly proportional to $\nnz(\hat{W})$,
the number of nonzero elements of $\hat{W}$. The total
computational 
cost of SRE is $O(c_{\sre}k_{\svd}\nnz(\hat{W}))$, where 
$c_{\sre}$ the the level of recursion and $k_{\svd}$ is the
number of Lanczos iteration steps. In general, $k_{\svd}$ depends on
the singular value gaps of $\hat{W}$. Also notice that
$\nnz(\hat{W})= n_w n$, where $n_w$ is the average number of
terms per document and $n$ is the total number of document.
Therefore, the total cost of SRE is in general linear in the
number of documents to be clustered.

\section{Connections to correspondence analysis}\label{se:corr}
In its basic form correspondence analysis is applied to an 
$m$-by-$n$ two-way
table of counts $W$ \cite{benz:92,gree:93,veri:99}. Let $w=\s(W)$,
the sum of all the elements of $W$, $D_X$ and $D_Y$ be diagonal
matrices defined in section \ref{se:svd}. Correspondence analysis
seeks to compute the largest singular triplets of the matrix 
$Z=(z_{ij}) \in \R^{m \times n}$ with
\[ z_{ij}= \frac{w_{ij}/w - (D_X(i,i)/w)(D_Y(j,j)/w)}
             {\sqrt{(D_X(i,i)/w)(D_Y(j,j)/w)}}.\]
The matrix $Z$ can be considered as the correlation matrix of two
group indicator matrices for the original $W$ \cite{veri:99}. 
We now show that the SVD of $Z$ is closely related to the
SVD of $\hat{W} \equiv D^{-1/2}_XWD_Y^{-1/2}$. 
In fact, in section \ref{se:svd},
we showed that $D_X^{1/2}e$ and $D_Y^{1/2}e$ are the left and right
singular vectors of $\hat{W}$ corresponding to the singular value one,
and it is also easy to show that all the singular values of
$\hat{W}$ are at most $1$. Therefore, 
the rest of the singular values and singular vectors of
$\hat{W}$ can be found by computing the SVD of 
the following  rank-one modification
of $\hat{W}$
\[D^{-1/2}_XWD_Y^{-1/2}-
\frac{D_X^{1/2}ee^TD_Y^{1/2}}{\|D_X^{1/2}e\|_2\|D_Y^{1/2}\|_2}\]
which has $(i,j)$ element
\[  \frac{w_{ij}}{\sqrt{D_X(i,i)D_Y(j,j)}} - 
     \frac{\sqrt{D_X(i,i)D_Y(j,j)}}{w} = w^2z_{ij},\]
and  is a constant multiple of the $(i,j)$ element of $Z$.
Therefore, normalized-cut based  cluster analysis and correspondence
analysis arrive at the same SVD problems even though they start with
completely different principles. It is worthwhile to explore
more deeply the interplay between these two different points of views and
approaches, for example, using the statistical analysis of
correspondence analysis to provide better strategy for selecting cut
points and estimating the number of clusters.

\section{Partitions with overlaps}\label{se:over}
So far in our discussion, we have only looked at {\it hard}
clustering, i.e., a data object belongs to one and only
one cluster. In many situations, especially when there are much
overlap among the clusters, it is more advantageous to allow
data objects to belong to different clusters. For
example, in document clustering, certain groups of words can
be shared by two clusters. Is it possible
to model this overlap using our bipartite graph model and also
find efficient approximate solutions? The answer seems to be yes,
but our results at this point are rather preliminary and we will
only illustrate the possibilities. Our basic idea is that when computing
$\Ncut(A,B)$, we should disregard the contributions of the
set of vertices that is in the overlap. More specifically,
let $X=A\cup O_X \cup \bar{A}$ and $Y=B\cup O_Y\cup \bar{B}$, where
$O_X$ denotes the overlap between 
the vertex subsets
$A\cup O_X$ and $\bar{A}\cup O_X$, and
$O_Y$ the overlap between $B\cup O_Y$ and $\bar{B}\cup O_Y$, we compute
\[ \Ncut(A,B,\bar{A},\bar{B}) =\frac{\cut(A,B)}{W(A,Y)+W(X,B)}\]\[
+\frac{\cut(\bar{A},\bar{B})}{W(\bar{A},Y)+W(X,\bar{B})}.\]
However, we can make 
$\Ncut(A,B,\bar{A},\bar{B})$ smaller simply by putting more
vertices in the overlap. Therefore, we need to balance these
two competing quantities: the size of the overlap and the modified
normalized cut $\Ncut(A,B,\bar{A},\bar{B})$ by minimizing
\[ \Ncut(A,B,\bar{A},\bar{B}) + \alpha(|O_X| + |O_Y|),\]
where $\alpha$ is a regularization parameter. How to find an
efficient method for computing the (approximate) optimal
solution to the above minimization problem still needs to be
investigated. We close this section by presenting an illustrative
example showing that in some situations, the singular vectors
already automatically separating the overlap sets while giving
the coordinates for carrying out clustering.

\begin{figure}[t]
\centerline{
\mbox{\psfig{file=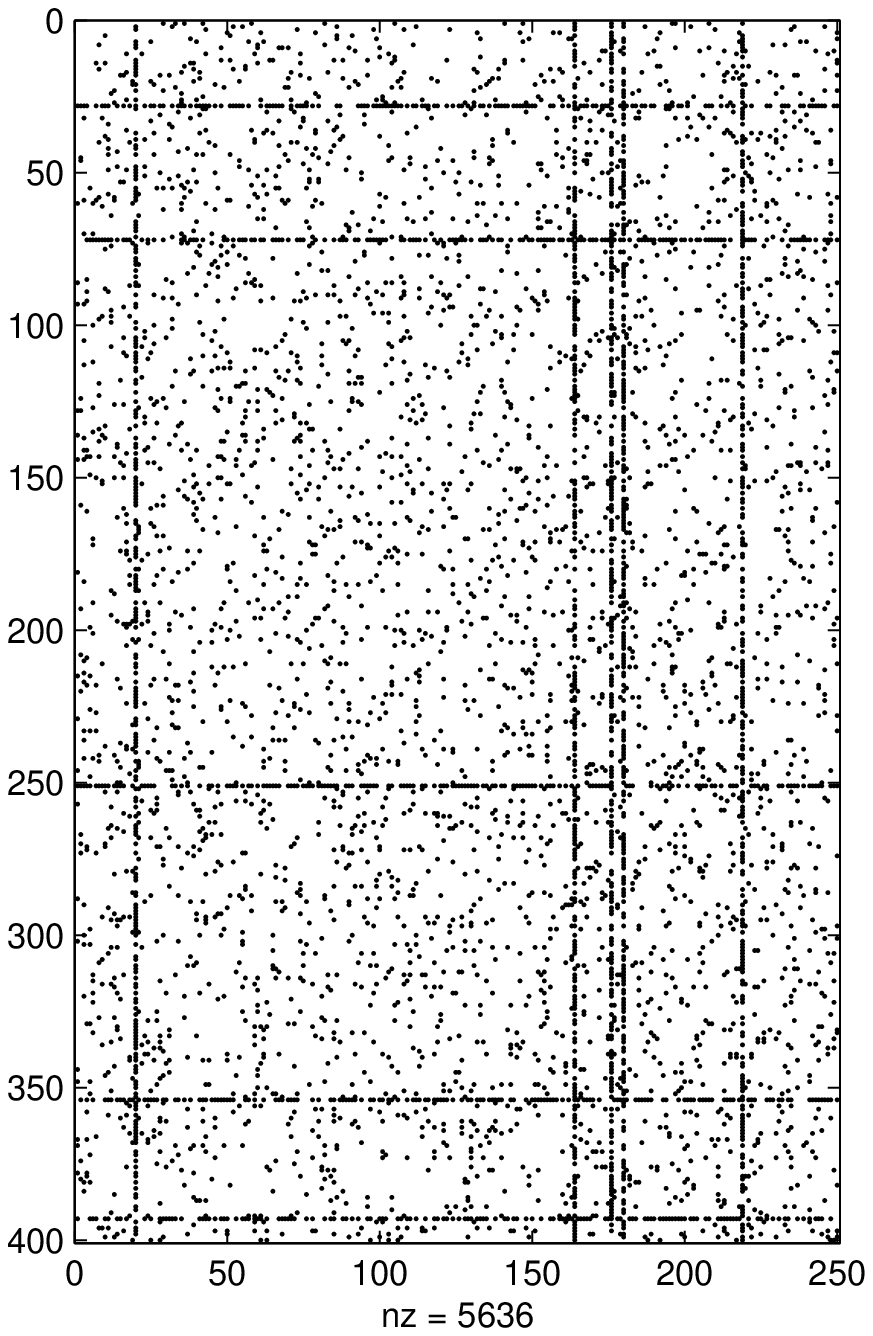,height=1.8in,width=1.6in}
\psfig{file=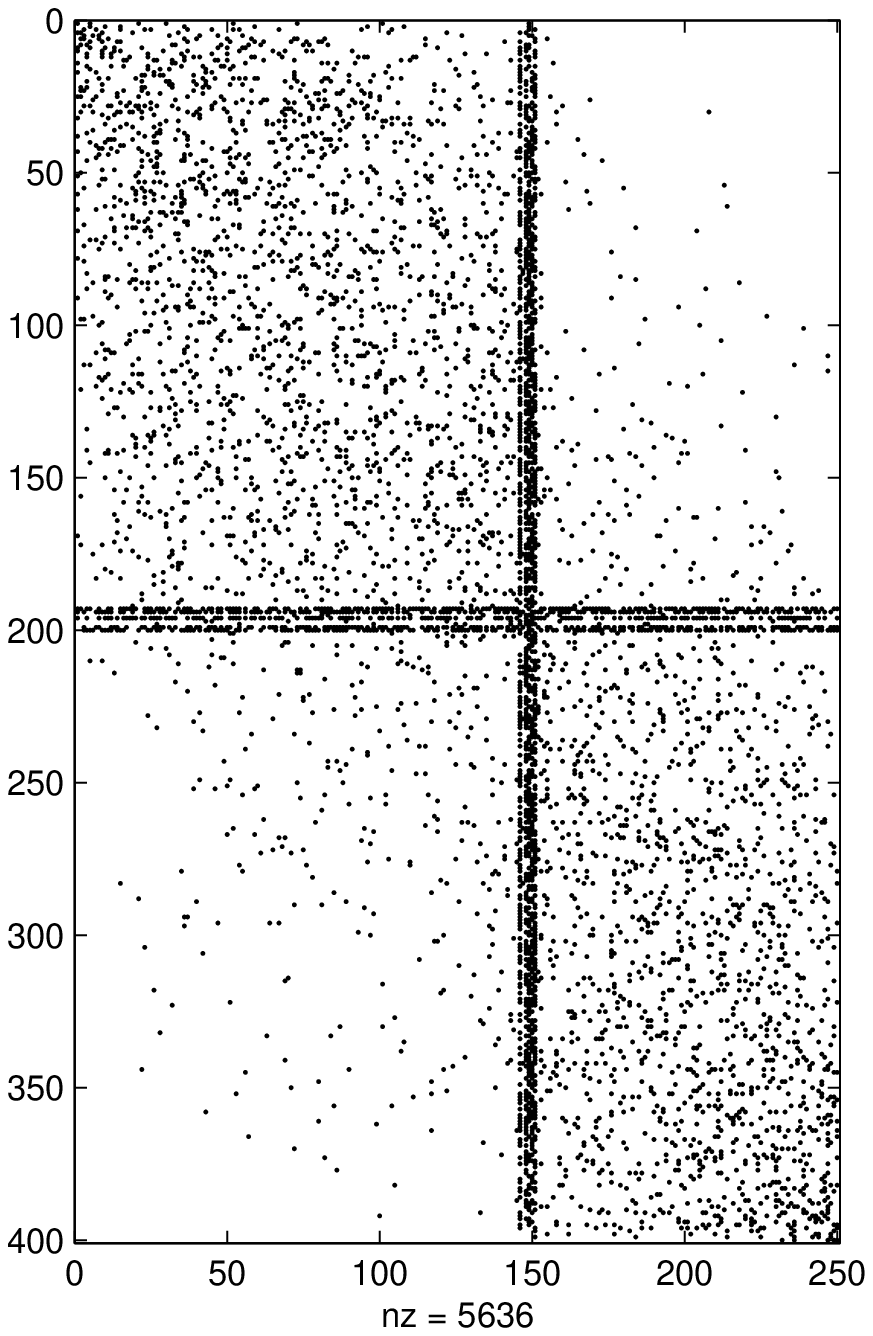,height=1.8in,width=1.6in}}}
\caption{Sparsity patterns of a test matrix before clustering
(left) and after clustering (right)}
\label{fi:op}
\end{figure}

{\sc Example 1.} We construct a sparse $m$-by-$n$ rectangular matrix 
\[ W = \left[\begin{array}{cc}
             W_{11}&W_{12}\\
             W_{21}&W_{22}
       \end{array}\right].\]
so that $W_{11}$ and $W_{22}$ are relatively denser than $W_{12}$
and $W_{21}$. We also add some dense rows and columns to the matrix $W$
to represent row and column overlaps.
The left panel of Figure \ref{fi:op} shows the sparsity pattern of 
$\bar{W}$,
a matrix obtained by randomly permuting
the rows and columns of $W$. We then compute the
second largest left and right singular vectors of 
$D_X^{-1/2} \bar{W} D_Y^{-1/2}$, say $x$ and $y$, then sort the rows and
columns of $\bar{W}$ according to the values of the entries in
$D_X^{-1/2}x$ and $D_Y^{-1/2}y$, respectively. The sparsity
pattern of this permuted $\bar{W}$ is shown on the right panel of
Figure \ref{fi:op}. As can be seen that the singular vectors not
only do the job of clustering but at the same time also
concentrate the dense rows and columns at the boundary of the two
clusters.

\section{Experiments}\label{se:exp}
In this section we present our experimental results on clustering
a dataset of newsgroup articles submitted to  20 
newsgroups.\footnote{
The newsgroup dataset together with the {\tt bow} toolkit for
processing it  
can be downloaded from
{\tt http://www.cs.cmu.edu/afs/cs/project/theo-11/www/}

\noindent
{\tt naive-bayes.html}.}
This dataset contains about
20,000 articles (email messages) evenly divided among the 20
newsgroups. We list the names of the newsgroups together
with the associated group labels (the labels will be
used in the sequel to identify the newsgroups).

\begin{verbatim}
       NG1: alt.atheism   
       NG2: comp.graphics 
       NG3: comp.os.ms-windows.misc   
       NG4: comp.sys.ibm.pc.hardware 
       NG5:comp.sys.mac.hardware   
       NG6: comp.windows.x 
       NG7:misc.forsale   
       NG8: rec.autos 
       NG9:rec.motorcycles   
       NG10: rec.sport.baseball 
       NG11:rec.sport.hockey 
       NG12: sci.crypt 
       NG13:sci.electronics   
       NG14: sci.med 
       NG15:sci.space   
       NG16: soc.religion.christian 
       NG17:talk.politics.guns   
       NG18: talk.politics.mideast 
       NG19:talk.politics.misc   
       NG20: talk.religion.misc 
\end{verbatim}

\begin{table*}\label{tb:12}
\caption{Comparison of spectral embedding (SRE), PDDP, and K-means
(NG1/NG2)
}
\begin{center}
\begin{tabular}{llrrr}
Mixture & SRE & PDDP & K-means \\ \hline\hline
50/50   & $92.12\pm 3.52$\% & $91.90\pm 3.19$\% $(53,10,37)$&$ 76.93\pm 14.42$\% $(82,2,10)$\\ \hline
50/100  & $90.57\pm 3.11$\% & $86.11\pm 3.94$\% $(86, 5,9)$&$ 76.74\pm 14.01$\% $(80,2,18)$\\ \hline
50/150  & $88.04\pm 3.90$\% & $78.60\pm 5.03$\% $(98, 0, 2)$&$68.80\pm 13.55$\% $(88, 0, 12)$\\ \hline
50/200  & $82.77\pm 5.24$\% & $70.43\pm 6.04$\% $(97,0,3)$&$69.22\pm 12.34$\% $(83,1,16)$\\ \hline
\end{tabular}
\end{center}
\end{table*}

\begin{table*}\label{tb:1011}
\caption{Comparison of spectral embedding (SRE), PDDP, and K-means
(NG10/NG11)
}
\begin{center}
\begin{tabular}{llrrr}
Mixture & SRE & PDDP & K-means \\ \hline\hline
50/50   & $74.56\pm 8.93$\% & $73.40\pm 10.07$\% $(56,6,38)$ &$61.61\pm 8.77$\% $(86,0,14)$\\ \hline
50/100  & $67.13\pm 7.17$\% & $67.10\pm 10.20$\% $(52,1,47)$ &$64.40\pm 9.37$\% $(59,1,40)$\\ \hline
50/150  & $58.30\pm 5.99$\% & $58.72\pm 7.48$\% $(52,1,47)$ &$62.53\pm 8.20$\% $(36,1,63)$\\ \hline
50/200  & $57.55\pm 5.69$\% & $56.63\pm 4.84$\% $(58,1,41)$ &$60.82\pm 7.54$\% $(39,2,59)$\\ \hline
\end{tabular}
\end{center}
\end{table*}

We used the {\it bow} toolkit to construct the term-document
matrix for this dataset, specifically we use the tokenization option
so that the UseNet  headers are stripped, and we also applied stemming
\cite{mcca:96}. Some of the newsgroups have large overlaps, for
example, the five newsgroups {\tt comp.* } about
computers. In fact several articles are posted to multiple newsgroups.
Before we apply clustering algorithms to the dataset, several
preprocessing steps need to be considered. Two standard steps
are weighting and feature selection. For weighting, we considered
a variant of tf.idf weighting scheme,
$\tf\log_2(n/\df),$
where $\tf$ is the term frequency and $\df$ is the document
frequency and several other variations
listed in \cite{bele:00}. 
For feature selection, we looked at three approaches 1)
deleting terms that occur less than certain number of
times in the dataset; 2) deleting terms that 
occur in less than certain number of
documents in the dataset; 3) selecting terms according to mutual
information of terms and documents defined as
\[ I(y) = \sum_{x} p(x,y)\log(p(x,y)/(p(x)p(y)),\]
where $y$ represents a term and $x$ a document \cite{slti:00}.
In general we found out that the traditional tf.idf based
weighting schemes do not improve performance for SRE. One possible
explanation comes from the connection with correspondence analysis,
the raw frequencies are samples of co-occurrence probabilities,
and the pre- and post-multiplication by $D_X^{-1/2}$ and
$D_Y^{-1/2}$ in $D_X^{-1/2}(D-W)D_Y^{-1/2}$ {\it automatically}
taking into account of weighting. We did, however, found out that
trimming the raw frequencies can sometimes improve 
performance for SRE, especially for the anomalous cases where
some words can occur in certain documents an unusual number of times,
skewing the clustering process.

For the purpose of comparison, we consider two other clustering
methods: 1) K-means method \cite{Gordon}; 2) Principal direction
divisive partion (PDDP) method \cite{bole:98}. K-means method is
a widely used cluster analysis tool. The variant we used employs
the Euclidean distance when comparing the dissimilarity between
two documents. When applying K-means,
we {\it normalize} the length of each document so that it has
Euclidean length one. In essence, we use the cosine of the angle
between two document vectors when
measuring their similarity. We have also tried K-means without
document length normalization, the results are far worse and therefore
we will not report the corresponding results. Since K-means method is
an iterative method, we need to specify a stopping criterion. For
the variant we used, we compare the centroids between two
consecutive iterations, and stop when the difference is smaller
than a pre-defined tolerance.

PDDP is another clustering method that utilizes singular
vectors. It is based on the idea of principal component
analysis and has been shown to
outperform several standard clustering methods
such as hierarchical agglomerative algorithm \cite{bole:98}. 
First each document is considered as a 
multivariate data point. The set of document is normalized
to have unit Euclidean length and then centered, i,e., let
$W$ be the term-document matrix, and $w$ be the average of
the columns of $W$. Compute the largest singular value triplet
$\{u,\sigma,v\}$ of
$W-we^T$. Then split the set of documents based on their 
values of the $v=(v_i)$ vector: one simple scheme is to 
let those with
positive $v_i$ go into one cluster and those
with nonnegative $v_i$  inot another cluster. Then the
whole process is repeated on the term-document matrices of 
the two clusters, respectively. Although both our clustering
method SRE and PDDP
make use of the singular vectors of some versions of the
term-document matrices, they are derived from fundamentally
different principles. PDDP is a feature-based clustering method,
projecting all the data points to the one-dimensional subspace
spanned by the first principal axis; SRE is a similarity-based
clustering method, two co-occurring variables (terms and
documents in the context of document clustering) are
simultaneously clustered. Unlike SRE, PDDP does not
have a well-defined objective function
for minimization. It only partitions the columns of
the term-document matrices while SRE partitions both of its
rows and columns. This will have significant impact on the
computational costs.
PDDP, however, has an  advantage that it can be applied to
dataset with both positive  and negative values while SRE can only be
applied to datasets with nonnegative data values. 

\begin{table*}\label{tb:1819}
\caption{Comparison of spectral embedding (SRE), PDDP, and K-means
(NG18/NG19)
}
\begin{center}
\begin{tabular}{llrrr}
Mixture & SRE & PDDP & K-means \\ \hline\hline
50/50  & $73.66\pm 10.53$\% & $69.52\pm 12.83$\% $(65,12,32)$ & $62.25 \pm 9.94$\% $(82,1,17)$\\ \hline
50/100   & $67.23\pm 7.84$\% & $67.84\pm 7.30$\% $(46,5,49)$& $60.91\pm 7.92$\% $(65,13,32)$\\ \hline
50/150  & $65.83\pm 12.79$\% & $60.37\pm 9.85$\% $(53,3,44)$ &$63.32\pm 8.26$\% $(58,3,39)$\\ \hline
50/200  & $61.23\pm 9.88$\% & $60.76\pm 5.55$\% $(40,1,59)$ &$64.50\pm 7.58$\% $(34,0,66)$\\ \hline
\end{tabular}
\end{center}
\end{table*}

\begin{table*}\label{tb:con}
\caption{Confusion matrix for newsgroups $\{2, 9, 10, 15, 18\}$
}
\begin{center}
\begin{tabular}{|l||c|c|c|c|c|}
\hline\hline
 &mideast &graphics & space & baseball &  motorcycles \\ \hline\hline
cluster 1&  87&  0&   0&    2&   0\\ \hline
cluster 2&  7&   90&  7&    6&   7\\ \hline
cluster 3&  3&   9&   84&   1&   1\\ \hline
cluster 4&  0&   0&   1&    88&  0\\ \hline
cluster 5&  3&   1&   8&    3&   92\\ \hline\hline
\end{tabular}
\end{center}
\end{table*}

{\sc Example 2.} In this example, we examine binary clustering
with uneven clusters. We consider three pairs of newsgroups:
newsgroups 1 and 2 are well-separated, 10 and 11 are 
less well-separated and 18 and 19 have a lot of overlap.
We used document frequency as the feature
selection criterion and delete 
words that occur in less than $5$ documents in each datasets we
used. For both K-means and PDDP we apply tf.idf weighting together
with document length normalization so that each document vector
will have Euclidean norm one. For SRE we trim the raw frequency
so that the maximum is $10$.
For each newsgroup pair, we select four
types of mixture of articles from each newsgroup: $x/y$ indicates
that $x$ articles are from the first group and $y$ articles are
from the second group. The results are listed in Table
1 for groups 1 and 2, Table 2 for groups 10 and
11 and Table 3 for groups 18 and 19. We list
the means  and standard deviations for  100 random samples.
For PDDP and K-means we also include a triplet of numbers
which indicates how many of the 100 samples SRE performs better (the first 
number), the same (the second number) and worse (the third number) than
the corresponding methods (PDDP or K-means).
We should emphasize that
K-means method can only find local minimum, and the results 
depend on initial values and stopping criteria. This is also
reflected by the large standard deviations associated with
K-means method.
From the three
tests we can conclude that both SRE and PDDP outperform K-means
method. The performance of SRE and PDDP are similar in balanced
mixtures, but SRE is superior to PDDP in skewed mixtures.

{\sc Example 3.} In this example, we consider an easy multi-cluster case,
we examine five newsgroups $2, 9, 10, 15, 18$ which
was also considered in \cite{slti:00}. We sample 100
articles from each newsgroups, we use mutual information for
feature selection.
We use minimum normalized cut as cut point for each level
of the recursion.
For one sample, Table 4 gives the confusion matrix.
The accuracy for this sample is $88.2$\%. We also tested two
other samples with accuracy  $85.4$\%
and $81.2$\% 
which compare
favorably
with those obtained for three samples with
accuracy $59$\%, $58$\% and $53$\% reported in \cite{slti:00}.
In the following we also listed the top few words for
each clusters computed by mutual information.

\begin{verbatim}
Cluster 1:
 armenian israel arab palestinian peopl jew isra
 iran muslim kill turkis war greek iraqi adl call

Cluster 2: 
 imag file bit green gif mail graphic colour
 group version comput jpeg blue xv ftp ac uk list

Cluster 3: 
 univers space nasa theori system mission henri 
 moon cost sky launch orbit shuttl physic work 

Cluster 4: 
 clutch year game gant player team hirschbeck 
 basebal won hi lost ball defens base run win

Cluster 5: 
 bike dog lock ride don wave drive black
 articl write apr motorcycl ca turn dod insur
\end{verbatim}

\section{Conclusions and feature work}\label{se:con}
In this paper, we formulate a class of clustering problems as
bipartite graph partitioning problems, and we show that
efficient optimal solutions can be found by computing the
partial singular value decomposition of some scaled edge weight
matrices. However, we have also shown that there still remain
many challenging problems. One area that needs further investigation
is the selection of cut points and number of clusters using 
multiple left and right singular vectors, and
the possibility of adding local refinements to improve
clustering quality.\footnote{It will be
difficult to use local refinement for PDDP
because it does not have a global objective function
for minimization.} Another area is to find
efficient algorithms for handling overlapping clusters. Finally,
the treatment of missing data under our bipartite graph model
especially when we apply our spectral clustering methods to
the problem of data analysis of recommender systems also deserves 
further investigation.

\section{Acknowledgments}
The work of Hongyuan Zha and Xiaofeng He was supported in
part  by NSF grant CCR-9901986. The work of Xiaofeng He, 
Chris Ding and Horst Simon was supported in
part  by Department of Energy through an LBL LDRD fund.

\bibliographystyle{plain}
\bibliography{ref}

\appendix
\section{Some proofs} 
In this appendix we prove three
results: 1) All the
eigenvalues of $D^{-1/2}WD^{-1/2}$ has absolute value at
most $1$. Equivalently, we need to prove that the eigenvalues
of the generalized eigenvalue problem $Wx = \lambda Dx$
has absolute value at
most $1$. In fact let $x=(x_i)_{i=1}^n$ and let $i$ be such that 
$|x_i| = \max |x_j|$, then it follows from
\[ \lambda d_i x_i = \sum_{j=1}^n w_{ij} x_j\]
that
\[ |\lambda| \leq \sum_{j=1}^n w_{ij}/d_i = 1.\]

2) We prove that
\[ \sigma_{\max}(D_X^{-1/2}WD_Y^{-1/2})
=\max_{x \neq 0, y \neq 0} \frac{2 x^TWy}{x^TD_Xx + y^TD_Yy}.\]
Let $\hat{x}= D^{1/2}_Xx$ and $\hat{y}= D^{1/2}_Yy$, then
\begin{equation}\label{eq:ff}
\frac{2 x^TWy}{x^TD_Xx + y^TD_Yy} = 
 \frac{2 \hat{x}^TD_X^{-1/2}WD_Y^{-1/2}\hat{y}}
{\hat{x}^T\hat{x} + \hat{y}^T\hat{y}}.\end{equation}
Let $D_X^{-1/2}WD_Y^{-1/2}=U\Sigma V^T$ be its SVD with
\[ U = [u_1, \dots, u_m], \quad V=[v_1,\dots, v_n] \]
and
\[ \Sigma = \diag(\sigma_1, \dots, \sigma_{\min\{m,n\}}), \quad
\sigma_1 = \sigma_{\max}(D_X^{-1/2}WD_Y^{-1/2}).\] Then
we can expand $\hat{x}$ and $\hat{y}$ as
\begin{equation}\label{eq:hh}
 \hat{x} = \sum_{i} \hat{x}_i u_i, \quad \hat{y} = \sum_{i} \hat{y}_i v_i, 
\end{equation}
and (\ref{eq:ff}) becomes
\[ \frac{2\sum_{i} \sigma_i \hat{x}_i\hat{y}_i}{\sum_i \hat{x}_i^2 + 
\sum_i \hat{y}_i^2}
\leq \frac{2\sigma_1 \sqrt{\sum_i \hat{x}_i^2}\sqrt{\sum_i \hat{y}_i^2}}
{\sum_i \hat{x}_i^2 + \sum_i \hat{y}_i^2} \leq \sigma_1.\]
Taking $\hat{x}_1=1$ and $\hat{y}_1=1$ achieves the maximum. 

3) Now we consider
the constraint
\[ x^TD_Xe + y^TD_Ye = 0\]
which is equivalent to 
$\hat{x}_1+\hat{y}_1=0$ using the expansions in (\ref{eq:hh}).
We can always scale the vectors $\hat{x}$ and $\hat{y}$ 
without changing the maximum so that
$\hat{x}_1 \geq 0$ and $\hat{y}_1 \geq 0$. 
Hence $\hat{x}_1+\hat{y}_1=0$ implies that
$\hat{x}_1=\hat{y}_1=0$. It is then easy to see that
\[\sigma_2 = \max\left\{ \frac{2x^TWy}{x^TD_Xx + y^TD_Yy}\;\; | \;\;
x^TD_Xe + y^TD_Ye = 0 \right\},\]
and the maximum is achieved by the second largest left and
right singular vectors of $D_X^{-1/2} W D_Y^{-1/2}$.

\end{document}